\documentclass[twocolumn,longbibliography,aps,prx,superscriptaddress]{revtex4-1}
\usepackage{graphicx}
\usepackage{dsfont}
\usepackage{subcaption}
\usepackage{natbib}
\usepackage[dvipsnames]{xcolor}
\usepackage{amsmath}
\usepackage{amsthm}
\usepackage{color}
\usepackage{amsfonts}
\newcommand{\vac}{|{\rm vac}\rangle}
\newcommand{\vacl}{|{\rm vac}\rangle \langle{\rm vac}|}
\newcommand{\ket}[1]{\left| #1 \right\rangle}
\newcommand{\bra}[1]{\left\langle #1 \right|}
\newcommand{\braket}[2]{\left\langle #1 | #2 \right\rangle}

\newcommand{\Proj}[1]{| #1\rangle\!\langle #1 |}

\newcommand{\eea}{\end{eqnarray}}
\newcommand{\bea}{\begin{eqnarray}}
\newcommand{\ee}{\end{equation}}
\newcommand{\be}{\begin{equation}}

\newcommand{\sch}{Schr\"{o}dinger }
\newcommand{\lettersection}[1]{\emph{#1.---}}
\newcommand{\appropto}{\mathrel{\vcenter{
  \offinterlineskip\halign{\hfil$##$\cr
    \propto\cr\noalign{\kern2pt}\sim\cr\noalign{\kern-2pt}}}}}
    \newcommand{\Log}{{\rm Log}}
    
    \newcommand{\idh}{\hat{\mathds{1}}}
    
\begin{document}
%\title{Efficient detection of an arbitrary single-photon state}
\title{How to Project onto an Arbitrary Single-Photon Wavepacket}

\author{Tzula B. Propp}
\email{tzpropp@unm.edu}
\affiliation{Department of Physics and
Oregon Center for Optical, Molecular \& Quantum Sciences\\
University of Oregon, Eugene, Oregon 97403}
\affiliation{Department of Physics and
Astronomy, Center for Quantum Information and Control (CQuIC), University of New Mexico, Albuquerque, New Mexico 87131}

\author{Steven J. van Enk}
\affiliation{Department of Physics and
Oregon Center for Optical, Molecular \& Quantum Sciences\\
University of Oregon, Eugene, Oregon 97403}

\begin{abstract}
The time-frequency degree of freedom of the electromagnetic field is the final frontier for single-photon measurements.
The temporal and spectral distribution a measurement retrodicts  (that is, the state it projects onto) is determined by the detector's intrinsic resonance structure. In this paper, we construct ideal and more realistic positive operator-valued measures (POVMs) that project onto arbitrary  single-photon wavepackets with high efficiency and low noise. We discuss applications to super-resolved measurements and quantum communication. In doing so we will give a fully quantum description of the entire photo detection process, give prescriptions for (in principle) performing single-shot Heisenberg-limited time-frequency measurements of single photons, and discuss fundamental limits and trade-offs inherent to single-photon detection.
 
\end{abstract}

\maketitle

\section{POVMs for Photo Detection}

Photo detection is at its core an information theoretic process; a measurement outcome---a click---reveals information about the outside world quantifiable in bits \cite{shannon1948}. In the case of a single-photon detector (SPD), a click is correlated (imperfectly) with the presence of a particular type of photon, thus revealing information about the presence of photons of that type along with whatever else in the world such a photon is correlated with. The most general quantum description of this process is in terms of a positive operator-valued measure (POVM), a set of positive operators $\hat{\Pi}_k$ that sum to the identity, where each $k$  corresponds to a different measurement outcome.  Given an arbitrary input state $\hat{\rho}$ the probability to obtain outcome $k$ is given by the Born Rule
\be
Pr(k)={\rm Tr}(\hat{\rho}\hat{\Pi}_k).
\ee
Generically, each POVM element $\hat{\Pi_k}$ can be writen as a weighted sum over orthonormal quantum states
\bea \hat{\Pi}_k = \sum_i w_{i}^{(k)} \ket{\phi_i^{(k)}}\bra{\phi_i^{(k)}} \label{POVM1}
\eea reducing to an ideal Von Neumann measurement only when the sum contains a single term with its weight $w^{(k)}$ equal to $1$ \cite{von1932}. 
The weight $w_i^{(k)}$ equals the conditional probability to obtain measurement outcome $k$ given input $i$. The \emph{posterior} conditional probability that, given an outcome $k$, we project onto input $i$ is given by Bayes' theorem  \cite{Bayes63}
\bea Pr(i|k)= \frac{w_{i}^{(k)}  Pr(i)}{Pr(k)} \label{bayes}
\eea 
with $Pr(k)=\sum_i w_{i}^{(k)}  Pr(i)$ and $Pr(i)$ the \emph{a priori} probabilities to get outcome $k$ and for input $i$ to be present, respectively \cite{kraus1983}. Through Bayes' theorem, an experimentalist is able to \emph{retrodict}---that is, update their probability distribution over possible inputs---but only if they know what measurement their detector actually performs.

Knowledge of the POVM is essential for both gaining information from a measurement device and characterizing detector performance, hence the experimental need for detector tomography \cite{luis1999,goltsman2005,coldenstrodt2009,lundeen2009,dauria2011,brida2012}. Commercial photo detectors are characterized by industry-standard figures of merit \cite{hadfield2009}, which can be calculated from a POVM (for an in-depth review, see Ref.~\cite{vanenk2017}). Here we will concern ourselves mostly with two figures of merit, detection efficiency and time-frequency uncertainty:

\lettersection{Efficiency}The maximum efficiency with which an SPD outcome $k$ (for instance, a single click) can be triggered by input single photon states is the maximum relative weight in (\ref{POVM1}) $\eta_{\max} = {\rm Max}_{i} [w_{i}^{(k)} ] $. The maximum efficiency  is achieved only when the input quantum state is one the measurement projects onto. This follows directly from the Born rule; $Pr(k|i) = {\rm Tr} [ \hat{\Pi}_k \hat{\rho}] \rightarrow w_{i}^{(k)}  $ if and only if $\hat{\rho}=\ket{\phi_i^{(k)}}\bra{\phi_i^{(k)}} $. 

\lettersection{Time-Frequency Uncertainty}The spectral uncertainty and (input-independent) timing jitter are determined entirely by the spectral and temporal widths of the states projected onto by the measurement outcome $k$ \cite{helstrom1974}, which form a retrodictive probability distribution. For any continuous variable $X$ (here either time $t$ or frequency $\omega$), we find it less convenient to use the variance as measure of uncertainty and instead define the uncertainty entropically \cite{helstrom1974,epi1975,oppenheim2010,vanenk2017,wild2019} 
\bea\label{uncert}
\Delta X^{(k)} = 2^{H^{(k)}_X} \delta X.
\eea Here $H^{(k)}_X$ is the Shannon entropy defined as
\bea\label{entropy} 
H_X^{(k)} =-\sum_j p(j|k) {\rm log}_2 p(j|k)
\eea with the sum over discretized $X$-bins of size $\delta X$. $p(j|k)$ is the \emph{a posteriori} probability for the detected photon to be in bin $j$ given outcome $k$, and is calculated 
 \bea p(j|k) = \int_{(j-1)\delta X}^{j\delta X} dX \sum_i Pr(i|k) | \phi_i(X) |^2
 \label{postprob}
\eea where we have defined a normalized distribution over $X$ given by the norm squared of the quantum state $ | \phi_i(X) |^2$ (where $\phi_i(X) \equiv \braket{X}{\phi_i}$). The conditional probability $Pr(i|k) $ is precisely the one from Bayes theorem (\ref{bayes}); $Pr(i|k)$ reduces to $w_i^{(k)} / \Omega^{(k)}$ in the case of a uniform prior \footnote{For inclusion of priors in updating information about the quantum state, see Ref.~\cite{Holevo}.}, where $\Omega^{(k)} = \sum_i w_i^{(k)} $ is the bandwidth \cite{vanenk2017}. Critically, $\Delta X^{(k)}$ is independent of the bin size $\delta X$ in the small-bin limit, even though the entropy $H_X^{(k)}$ is strongly dependent on the bin size. One can verify that this definition of uncertainty yields a Fourier time-frequency uncertainty relation   \cite{epi1975}
\be
\Delta\omega\Delta t \geq e\pi.
\ee

In this paper, we construct  POVMs capable of projecting onto arbitrary quantum states (including minimum-uncertainty Gaussian wavepackets) with high efficiency. The construction of measurements projecting onto arbitrary single-photon states is critical in quantum optical and quantum communication experiments. Mismatch between the single-photon state generated and the state projected onto by the measurement induces an irreversible degradation in detection efficiency. As we show in section II, arbitrary quantum state measurements can be accomplished using a simple time-dependent two-level system. In section III, we show that realistic implementations of minimum-uncertainty POVMs projecting onto arbitrary quantum states are possible, even after including the effects of the initial coupling of the photo detecting system to the external world (transmission), the conversion of a single excitation into a macroscopic signal (amplification), and a noisy classical measurement of that final signal. The capacity to efficiently project onto (a small set of) orthogonal single-photon states enables a wide range of quantum information and quantum optical applications, as we discuss in section IV. In section V, we will discuss the fundamental limits and tradeoffs to photo detector performance that  manifest in our work, as well as experimental implementation of the derived POVM (for instance, using a lens to focus light onto a molecule whose state is monitored with electron shelving \cite{dehmelt1975}) in detail. From a foundational perspective, a procedure to build measurements efficiently projecting onto minimum-uncertainty Gaussian single-photon wavepackets paves the way for future tests of fundamental quantum theory.

\section{Simplified Measurements Projecting Onto Arbitrary Single-Photon States}

We will now discuss how to construct a simple POVM that efficiently projects onto an arbitrary single-photon wavepacket. To aid us, we will now make four simplifying assumptions. First, we will consider only the time-frequency degree of freedom of the electromagnetic field, as the other degrees of freedom (e.g. polarization) can be efficiently sorted prior to detection in a pre-filtering process \cite{osullivan2012,Bouchard2018,Fontaine2019}. Second, we consider only a single excitation incident to the photo detector. Multiple photons can always be efficiently multiplexed to achieve a photon number resolution using SPD pixels \cite{Nehra20}. Third, we will not model a continuous measurement (as briefly discussed in the appendix of \cite{proppnet}), but instead a discretized measurement where at a particular time $T$ we ascertain whether or not a photon has interacted with the SPD, ending the measurement. Lastly, we will consider only a binary-outcome photo detector, ``click'' or ``no click.'' This simplifies the POVM so that it only contains the two elements $\hat{\Pi}_T$ and $\hat{\Pi}_{0}$, both projecting onto the Hilbert space of  single photon states and the vacuum state. Generalizations to non-binary-outcome SPDs are straightforward: one can concatenate binary-outcome POVMs to generate non-binary-outcome experiments.

We now begin construction of the POVM $\{\hat{\Pi}_T,\,\hat{\Pi}_0\}$ in earnest. Consider a two-level system with time-dependent transition frequency $\Delta(t)$, with time-dependent coupling to a Markovian external electromagnetic continua of states \footnote{Non-markovianity of the external continua can be included via couplings to fictitious discrete states or pseudomodes, see Refs.~\cite{pseudomodes2,pseudomodes3,pseudomodes4,pseudomodes1,pseudomodes5}.}. Experimentally, a time-dependent decay rate $\kappa(t)$ is induced by a rapid variation of density of states \cite{Mart2004,Thyrrestrups2013} and a time-dependent resonance $\Delta(t)$ can be varied with a time-dependent external electric field (Stark effect, \cite{Stark1914}) or through a two-channel Raman transition \cite{Eberly1996}.

\begin{figure*}[t!]
        \begin{subfigure}[b]{0.475\textwidth}
            \centering
            \includegraphics[width=\textwidth]{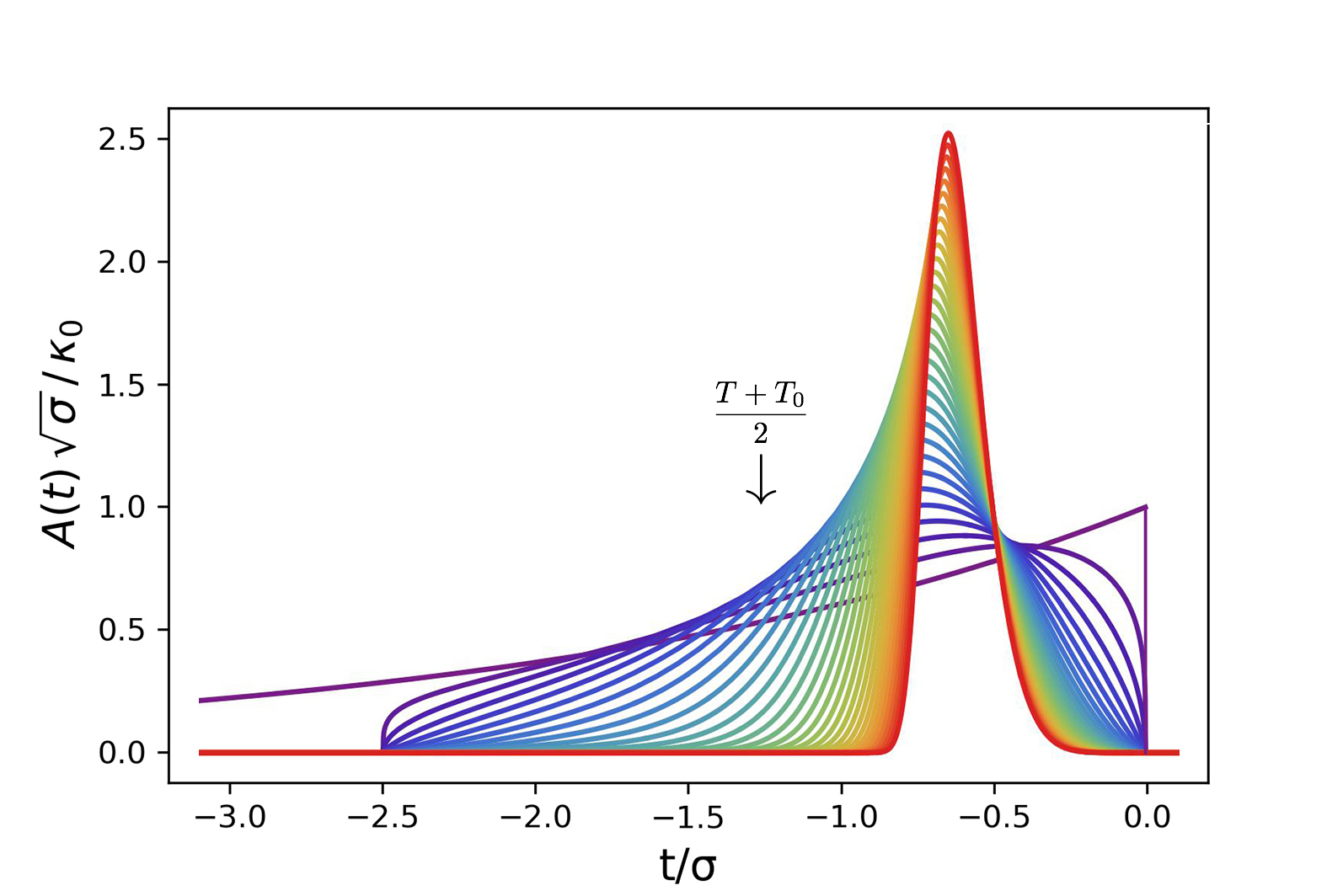}
            \caption[]%
            {{\small }}    
            \label{rainbow2psi}
        \end{subfigure}
        \begin{subfigure}[b]{0.475\textwidth}  
            \centering 
            \includegraphics[width=\textwidth]{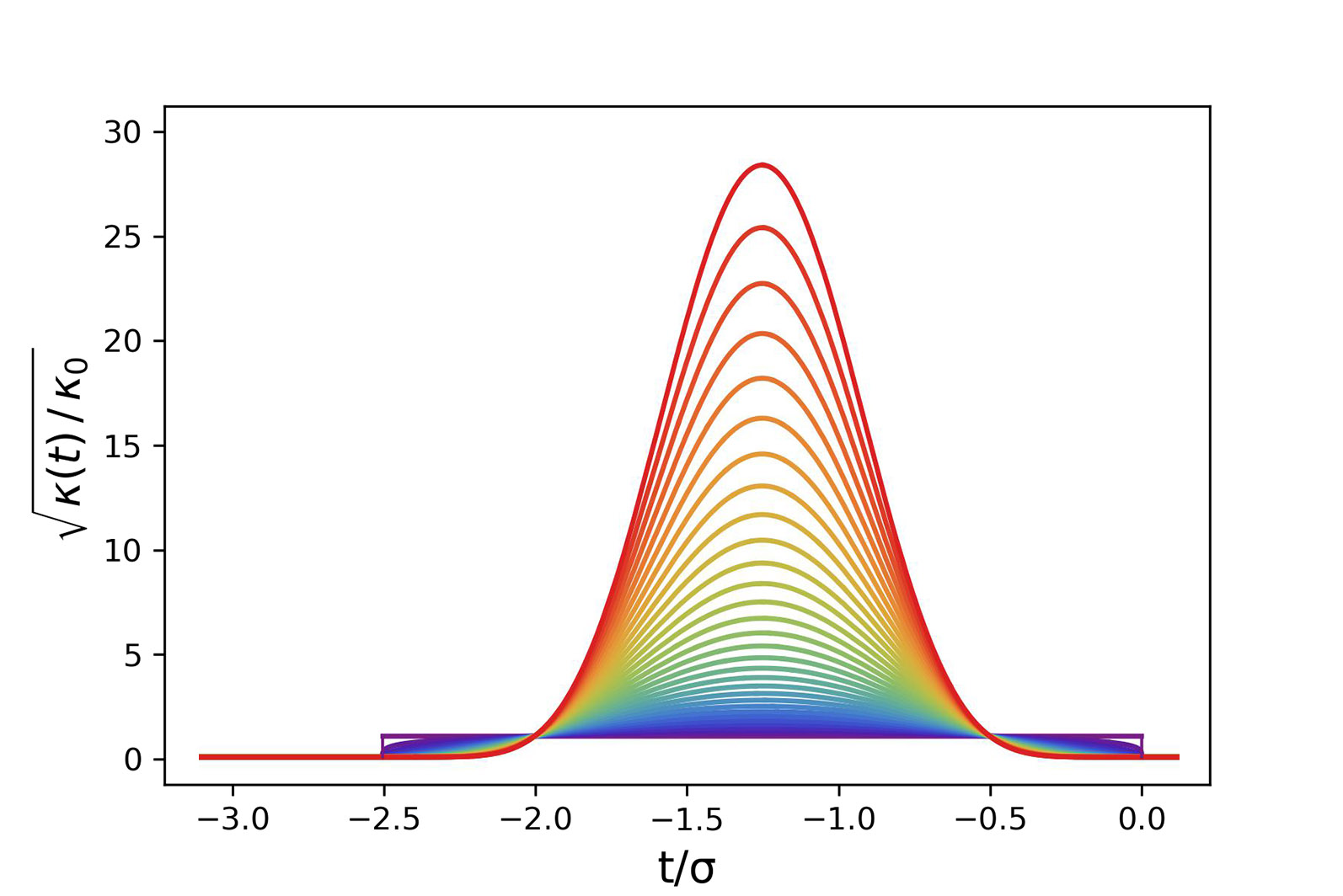}
            \caption[]%
            {{\small }}    
            \label{rainbow2kappa}
        \end{subfigure}
        \caption{\small (a) Retrodictive probability amplitudes $A(t)=|\Psi(t)|$ defined in (\ref{timeCSolPack}) are plotted for polynomial decays of the form $\kappa(t) =\kappa_0 \left(\frac{t-T_0}{\sigma}\right)^n\left(\frac{T-t}{\sigma}\right)^n$ for $T_0\leq t\leq T$, with the polynomial order $n$ varying from $0$ (violet, lowest peak) to $15$ (red, highest peak) denoted by the color and time measured w.r.t. detection time $T$. (b) The square root of the corresponding decay rates $\sqrt{\kappa(t)}$ are plotted themselves, with polynomial order varying from $0$ (violet, bottom within the central region) to $15$ (red, top within the central region). Except for $n=0$ (also in Fig. \ref{rainbow} and included here for reference), these decays continuously become non-zero at $T_0=-2.5\sigma$ (when the photo detector is turned on) so that ${\cal W}$ from (\ref{calW}) is strictly less than unity. In (a), we observe that the time of maximum coupling strength $t=\frac{T+T_0}{2}$ precedes the time of maximum retrodictive probability amplitude. For no order $n$ are the retrodictive probability distributions continuously differentiable; measurements with polynomial couplings do not project onto onto smooth wavepackets unlike the couplings plotted in Fig. \ref{dataCouplingSimp} and Fig. \ref{OrthogonalState}.} 
        \label{rainbow2}
\end{figure*}

In the quantum trajectory picture we can assume the state of the two-level system is pure and there are two types of evolution of $\ket{\psi(t)} $: Schr\"{o}dinger-like smooth evolution with a non-Hermitian effective Hamiltonian and quantum jumps (at random times) \cite{gardiner1992,carmichael1993}. Then the time-dependent (unnormalized) state of the two-level system is written $\ket{\psi(t)} = C_0 (t) \ket{0} + C_1 (t) \ket{1}$.  A quantum jump will always correspond to the excitation leaking out of the system and so, in the absence of a dark counts, we only need consider the Schr\"{o}dinger-like evolution in order to determine $\hat{\Pi}_T$. In this picture, the quantum state of the two-level system remains pure with the time-dependent excited state amplitude $C_1(t)$ obeying  a Langevin equation of the form \footnote{For similar treatments of universal quantum memory and, more recently, a quantum scatterer, see Refs.~\cite{lukin2007} and ~\cite{Molmer2019,Molmer2020}, respectively.}

\bea\label{simpleC}
\dot{C}_1(t)= -\frac{\kappa(t)}{2} C_1 (t) -i\Delta(t) C_1 (t) + \sqrt{\kappa(t)} f(t)
\eea where $f(t)$ is a normalized input photon wavepacket \cite{input1985, Gheri1998}. 
We can solve this equation with the result
\be
C_1(t)=\int_{T_0}^t dt'' f(t'')\sqrt{\kappa(t'')}\exp\left[-\int_{t''}^t dt' D(t')\right],
\ee
where
\be
D(t)=i\Delta(t)+\frac{\kappa(t)}{2},
\ee
and where $T_0$ is a time in the distant past where our photodetector was still off, so that $\kappa(T_0)=0$ and $C_1(T_0)=0$. Our measurement consists in checking if the system is in the excited state at time $t=T$.
The probability to obtain a positive result (corresponding to detecting the incident photon wavepacket)
is $|C_1(T)|^2$. We can write
\be
C_1(T)=\int_{T_0}^T dt \Psi^*(t)f(t) 
\ee
with 
\be\label{timeCSolPack}
\Psi^*(t)=\sqrt{\kappa(t)}\exp\left[-\int_{t}^T dt' D(t')\right].
\ee
Whereas $f(t)$ is a normalized wave function, $\Psi(t)$ is subnormalized for finite $T_0$, since
\be\label{calW}
{\cal W}=\int_{T_0}^T dt |\Psi(t)|^2=1-\exp
\left[-\int_{T_0}^T\! dt\, \kappa(t)\right].
\ee $\Psi(t)$ is a complex function with real amplitude  $A(t)$ and phase $\phi(t)$ such that $\Psi(t) = A(t) e^{-i \phi(t)}$. We can interpret $|\Psi(t)|^2$ as a retrodictive probability distribution over times $t$; indeed, the conditional probability that a photon entered the system between time $t$ and $t+dt$ given a detector ``click'' at a (later) time $T$ is $Pr(t|T)=|\Psi(t)|^2 dt$. 

We plot in Fig. \ref{rainbow2psi} the retrodictive probability distribution amplitudes $A(t)=|\Psi(t)|$ corresponding to simple polynomial decay rates $\kappa(t) =\kappa_0 \left(\frac{t-T_0}{\sigma}\right)^n\left(\frac{T-t}{\sigma}\right)^n$ for $T_0\leq t\leq T$, which are themselves plotted in Fig. \ref{rainbow2kappa}. For $n>0$, these polynomial decay rates incorporate finite response to detector on-time and off-time. We observe that the time of maximum likelihood is determined by competition between two effects, the probability of photo absorption and the rate at which the excited state amplitude decays, both of which are directly determined by $\kappa(t)$. This latter effect drives down the probability amplitude for the distant past. As a result, the peaks of the retrodictive distributions in Fig. \ref{rainbow2psi} do not match the peaks of the decay rates in Fig. \ref{rainbow2kappa} at time $\frac{T_0+T}{2}$ (when absorption is highest), but are instead located at a somewhat later time when the two effects balance out. Only a constant $\kappa(t)=\kappa_0$ ($n=0$) yields a non-zero probability amplitude at $t=T$ (where it is maximum).  For a constant decay rate, the most likely time that a photon entered the system is \emph{now} whereas for a time-dependent decay rate with $\kappa(T)=0$, there is some time of maximum likelihood determined by competition between the two effects.

\begin{figure*}[t!]
        \begin{subfigure}[b]{0.475\textwidth}
            \centering
            \includegraphics[width=\textwidth]{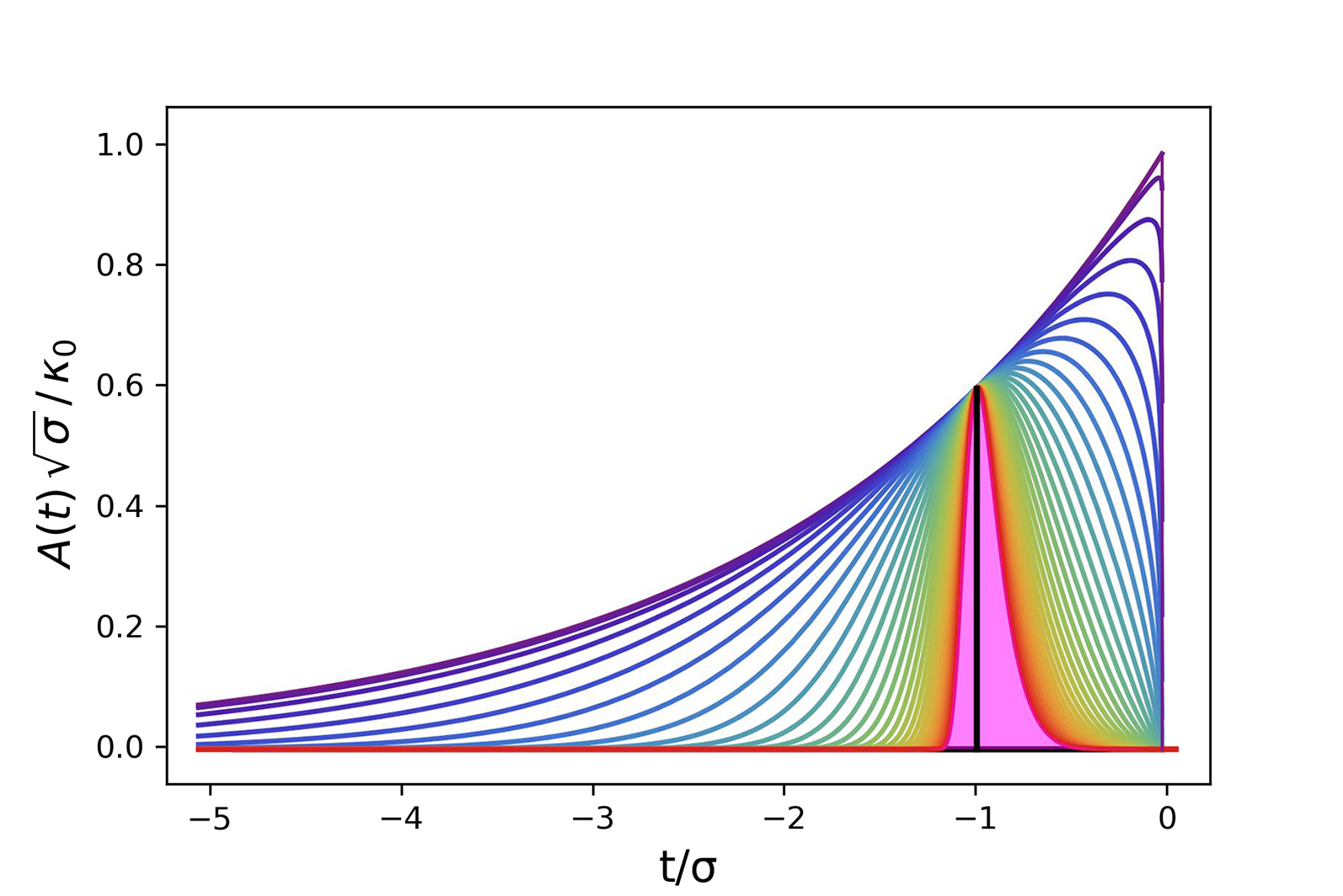} % rainbowcouplingnewermod.jpg
            \caption[]%
            {{\small }}    
            \label{rainbowpsi}
        \end{subfigure}
        \begin{subfigure}[b]{0.475\textwidth}  
            \centering 
            \includegraphics[width=\textwidth]{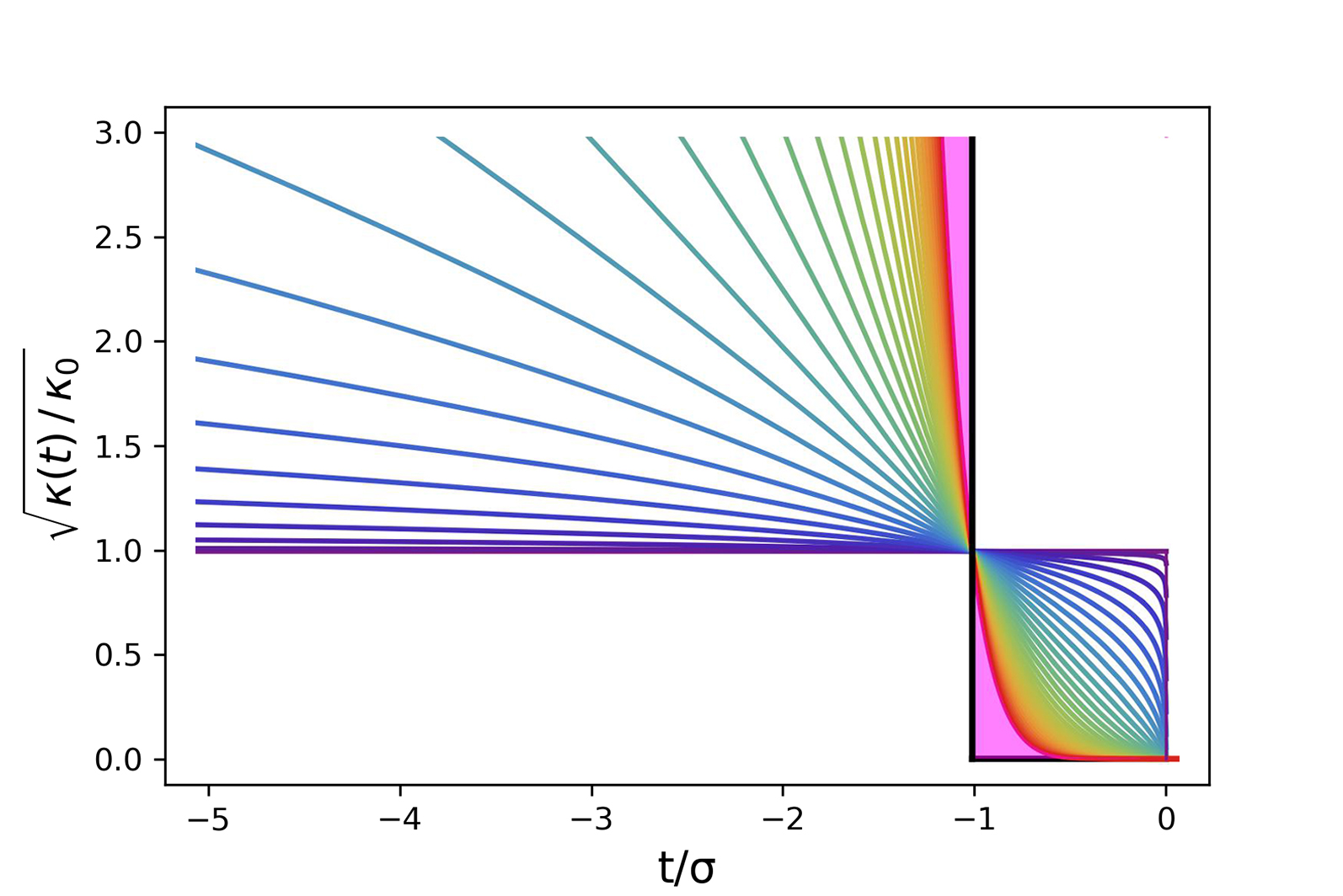}% rainbowkappa.jpg
            \caption[]%
            {{\small }}    
            \label{rainbowkappa}
        \end{subfigure}
        \caption{\small (a) The retrodictive probability amplitudes $A(t)=|\Psi(t)|$ defined in (\ref{timeCSolPack}) are plotted for polynomial decays of the form $\kappa(t) =\kappa_0 \left(\frac{T-t}{\sigma}\right)^n$ with polynomial order $n$ varying from $0$ (violet, top) to $15$ (red, bottom) denoted by color. (b) The square root of the corresponding decay rates $\sqrt{\kappa(t)}$ are plotted themselves, with polynomial order varying from $0$ (violet, top for $t/\sigma > -1$) to $15$ (red, bottom for $t/\sigma>-1$). Couplings and retrodictive probability amplitudes for higher polynomial order $n>15$ completely fill in the magenta shaded regions, approaching their asymptotic limits as $n\rightarrow \infty$ (black). In these plots, time is measured w.r.t. time of detection $T$ and we set $T_0=-\infty$ so that ${\cal W}=1$ (\ref{calW}). As in Fig. \ref{rainbow2}, for no order $n>0$ are the retrodictive probability distributions continuously differentiable, and the implemented measurements will not project onto smooth wavepackets unlike those described in Fig. \ref{dataCouplingSimp} and Fig. \ref{OrthogonalState}. } 
        \label{rainbow}
\end{figure*}

In Fig. \ref{rainbowpsi}, retrodictive probability distribution amplitudes are plotted for similar polynomial decays of the form $\kappa(t) =\kappa_0 \left(\frac{T-t}{\sigma}\right)^n$ taking into account finite off-time only, plotted in Fig. \ref{rainbowkappa}.  Although for $n>0$ the decay rate $\kappa(t)$ is not normalized and diverges as $t\rightarrow -\infty$, the retrodictive probability amplitudes $A(t)$ generated are still well-behaved. This is because $\kappa(t)$ directly corresponds to the decay rate of the excited state amplitude, so that a large $\kappa(t)$ drives down the probability amplitude for the distant past. As $n$ increases, this divergence in $\kappa(t)$ occurs faster so that for $t-T<\sigma$ the decay rate $\kappa(t)$ is very large, going suddenly to near-zero for $t-T> -\sigma$, completely filling in the shaded region and tending towards the black line for $n\rightarrow \infty$ in Fig. \ref{rainbowkappa}. These correspond to increasingly narrow retrodictive temporal distributions \footnote{A narrower temporal retrodictive distribution results in a broader Fourier-transform (spectral retrodictive distribution). Since the retrodictive distributions are not Gaussian, product-uncertainty will not be minimum.}, filling in the shaded region in Fig. \ref{rainbowpsi}. Since the decay rate $\kappa(t)$ is very large for $t-T< -\sigma$ an excitation absorbed before $t=-\sigma$ will leak back out, and since the coupling to the continuum $\sqrt{\frac{\kappa(t)}{2\pi}}$ is very small for $t-T> -\sigma$ excitations incident after $t=-\sigma$ will not be absorbed. In the limit $n\rightarrow \infty$ (depicted in Fig. \ref{rainbow} in black), the retrodictive probability distribution is zero everywhere except at $t=-\sigma$---the only time at which an excitation can enter the system and not immediately leak back out.

%
%
%  \begin{figure}[t] % "ht" = here or top
%	\includegraphics[width=1\linewidth]{rainbowcouplingnewermod.jpg} % For Mac OS X PDFs (else append .eps)
%	\caption{The retrodictive probability amplitudes $|\Psi(t)|$ defined in (\ref{timeCSolPack}) are plotted for polynomial decays of the form $\kappa(t) =\kappa_0 \left(\frac{T-t}{\sigma}\right)^n$ with polynomial order $n$ varying from $0$ (violet, top) to $40$ (red, bottom) denoted by color. Here time is measured w.r.t. time of detection $T$ and we set $T_0=-\infty$ so that ${\cal W}=1$ (\ref{calW}). For no order $n$ are the retrodictive probability distributions continuously differentiable; these simple polynomial couplings do not yield measurements projecting onto smooth wavepackets unlike the couplings plotted in Fig. \ref{dataCouplingSimp} and Fig. \ref{OrthogonalState}. Note that $\kappa(t)$ is not normalized and diverges as $t\rightarrow -\infty$, yet it results in well-behaved, normalized retrodictive probability amplitudes.}
%	\label{rainbow}
%\end{figure}

Having defined a retrodictive probability distribution in (\ref{timeCSolPack}), we can define a normalized single-photon state 
\bea\label{simpleState}
\ket{\Psi_T} ={\cal W}^{-1/2} \int_{T_0}^{T}\! dt\, \Psi(t) \hat{a}^\dagger_{\rm in} (t) \vac\nonumber \\
\eea with the creation operator $\hat{a}^\dagger_{\rm in} (t) $ acting on the input continuum of states. 
The arbitrary input single-photon state (which may have been created long before our detector was turned on at $T_0$ or long after the measurement ended at time $T$) is

\bea\label{inputStateDef}
\ket{f} &= \int_{-\infty}^{\infty} dt f(t) \hat{a}^\dagger_{\rm in} (t) \vac.
\eea
The commutator relation for the input field operator is $[\hat{a}_{\rm in} (t), \hat{a}^\dagger_{\rm in} (t')] = \delta(t-t')$.

The probability for an arbitrary  input photon wavepacket $f(t)$ to result in the system being found in the excited state at a time $T$ is 
 $|C_1(T)|^2=
	{\cal W}\braket{f}{\Psi_T}\braket{\Psi_T}{f} $.
We rewrite this probability in terms of a POVM element containing a single element 
 \bea\label{simpleAns}
\hat{\Pi}_T = {\cal W}\ket{\Psi_T}\bra{\Psi_T}
\eea 
The positive measurement outcome does not project onto times \emph{after} we have checked if the system is in the excited state, nor onto times before the detector was turned on.
(The other POVM element, describing the no-click outcome, does project onto all times.)

\begin{figure}[t] % "ht" = here or top
	\includegraphics[width=1\linewidth]{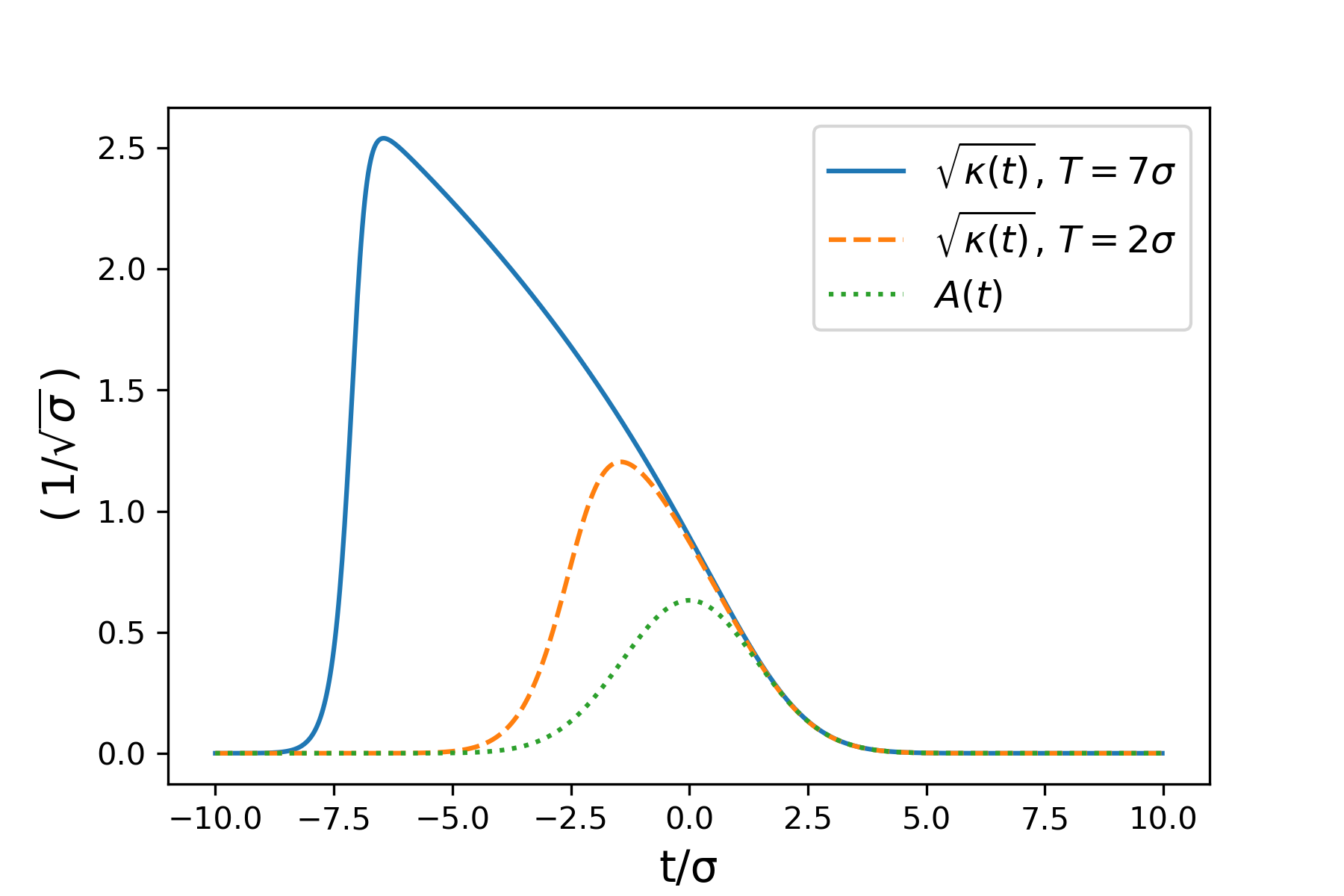} % For Mac OS X PDFs (else append .eps)
	\caption[]{The square root of the time-dependent couplings $\kappa(t)$ defined in (\ref{ErrfSol}) generating the minimum uncertainty (Gaussian) wavepacket are plotted for times of detection $T=7\sigma$ (blue solid line, ${\cal W} = 1-10^{-10}$) and $T=2\sigma$ (orange dashed line, ${\cal W} = 0.98$) in units of $1/\sqrt{\sigma}$ with detector on-time $T_0=-\infty$. The retrodictive probability amplitude is now a well-defined wavepacket, with amplitude $A(t)$ (green dotted line) plotted in the rotating frame (without the fast oscillations at the central frequency $\omega_0$), also in units of $1/\sqrt{\sigma}$. Time is measured w.r.t. the wavepacket's central time $t_0$. For times near $t\approx T$ the coupling $\sqrt{\kappa(t)}$ is approximately Gaussian as an excitation absorbed at this time will not have sufficient time to decay back out. For earlier times, $\sqrt{\kappa(t)}$ is strictly larger than $A(t)$ and is skewed towards earlier times to incorporate the time it takes the system to respond ($\sim 1/\kappa(t)$); one needs to prepare the two-level system for the photon, accomplished via a ``sharkfin'' coupling that precedes the single-photon wavepacket \cite{lukin2007}. Near $t=-T$, we observe the coupling $\sqrt{\kappa(t)}$ rapidly drops to zero, which is a direct consequence of the temporal wavepacket's symmetry about $t_0$. [From (\ref{timeCSolPack}) one can quickly verify that, if $A(t)$ is time-symmetric around $t=t_0$, the coupling $\kappa(t)$ must satisfy $\kappa(t_0-t) = e^{\int_{t_0-t}^{t_0+t} dt' \kappa(t')} \kappa(t_0+t)$.]}
	\label{dataCouplingSimp}
\end{figure}

To the extent our detector has been open long enough, such that ${\cal W}\rightarrow 1$, our detector could act as a perfectly efficient detector for a specific single-photon wavepacket with temporal mode function $\Psi(t)$ \footnote{For measurements projecting onto a Gaussian wavepacket as in (\ref{ErrfSol}) and Fig. \ref{dataCouplingSimp}, the weight (\ref{calW}) has the simple form ${\cal W} = \frac{1}{2} {\rm Erf}[\frac{t_0-T_0}{\sqrt{2}\sigma}] + \frac{1}{2} {\rm Erf}[\frac{T-t_0}{\sqrt{2}\sigma}]$, going to unity for $T,T_0\gg \sigma$.}. This wavepacket is the time reverse of the wavepacket that would be emitted by our two-level system if it started in the upper state $\ket{1}$ \footnote{An alternative to directly solving (\ref{simpleC}) is to find the Green's function $G(t)$ of the time-reversed problem: at $t_0$ the two-level system is started in the excited state and at time $T$ we check whether the excitation has leaked out. Taking $t\rightarrow T-t$, one arrives at (\ref{simpleState}) with $\Psi(t)=G(T-t)$. This approach is less direct but it does clarify the role of the Green's function: propagating back in time starting from $t = T$ (when the photon is detected) back to the infinite past, which indeed is what the POVM does as well (Fig. \ref{timeFlip}).}.

%This results in an alternative version of (\ref{simpleState}) 
%
%\bea\label{simpleStatetwo}
%\ket{G_T} &= \int_{-\infty}^{T} dt \sqrt{\kappa} G_{01}^*(T-t) \hat{a}^\dagger_{\rm in} (t) \vac \\
%&= \int_{-\infty}^{T} dt \sqrt{\kappa} G_{01}^*(t) \hat{a}^\dagger_{\rm in} (T-t) \vac
%\eea }

%\nonumber 

 %&= \int_{-\infty}^{T} dt \sqrt{\kappa} G_{01}^*(t) \hat{a}^\dagger_{\rm in} (T-t) \vac
%\eea with the creation operator $\hat{a}^\dagger_{\rm in} (t) $ acting on the input continuum of states. The Green's function in (\ref{simpleStatetwo}) describes how the system evolves from the ground state to the excited state. The second line of (\ref{simpleStatetwo}) shows one is really propagating back in time starting from $t = T$ (when the photon is detected) back to the infinite past.}%, which indeed is the role of the POVM as shown in Fig. \ref{timeFlip}

For this simple system, the POVM element is both pure (containing just one term  \footnote{We define purity of the POVM element ${\rm Pur}[\hat{Pi}_k] = \frac{{\rm Tr}[\hat{\Pi}_k^2]}{\left({\rm Tr}[\hat{\Pi}_k]\right)^2}\leq 1$ where the upper limit is reached only when the POVM element projects onto a single state.}) and (almost) maximally efficient (the weight ${\cal W}$ may approach unity as close as we wish).

Here we observe an obvious  trade-off between efficiency and photon counting rate: one cannot project onto a long single-photon wavepacket in a short time interval without cutting off the tails, lowering the overall detection efficiency \footnote{The limitation to photon counting rate imposed by efficient detection of long temporal wavepackets is avoided via signal multiplexing, see Ref.~\cite{Nehra20}.}. 

The two-level system described in Eq. (\ref{simpleC}) is a special case but an important one; the two-level system is often a very good approximation of more complicated systems near-resonance \footnote{For an arbitrary multi-level time-independent structure, we will end up with a system of equations governing discrete state evolution
\bea\label{generalC}
\dot{\vec{C}}(t)= \mathbf{M} \vec{C} (t) + \vec{S}(t)
\eea with $\mathbf{M}$ a time-independent matrix and $\vec{S}(t)$ a time-dependent
(inhomogeneous) source term describing the input photon. The solution is then always of the form
\bea\label{generalCSol}
\vec{C}(t)= e^{\mathbf{M} t} \vec{C} (t_0) + \int_{t_0}^t dt' e^{\mathbf{M} (t-t')} \vec{S}(t').
\eea 
Writing $e^{\mathbf{M} (t-t')}$ as a Green's matrix, we can identify elements that correspond to transitions to the final monitored discrete state (detector outcomes) through standard numerical techniques \cite{duffy2001}.}. In this paper, we will focus on the simple time-dependent system (\ref{simpleC}) as it is sufficiently general to perform a measurement described by any time-independent system, and more \footnote{In particular, time-independent systems cannot achieve Heisenberg-limited measurements of time and frequency. This is because networks of discrete states experience a natural spectral broadening that is Lorentzian. While Gaussian broadening can additionally occur (for instance, due to Doppler shifts in atomic distributions \cite{siegman86}) this only increases the product uncertainty further from the minimum of $\Delta \omega \Delta t = e\pi$ \cite{epi1975}, attained \emph{only} by pure measurements projecting onto Gaussian wavepackets.}. Indeed, (\ref{simpleC}) is general enough to project onto a \emph{completely arbitrary} single-photon wavepacket, a result we will now prove.

\lettersection{Proof} Consider a photon with complex wavepacket $\Psi^*(t)=A(t)e^{i\phi(t)}$, positive amplitude $A(t)$, and phase $\phi(t)$. Inserting this into (\ref{timeCSolPack}), we arrive at two separate expressions 

\bea\label{sepExpress}
A(t)&=& \sqrt{\kappa(t)} e^{-\int_t^T dt' \frac{\kappa(t')}{2}}\nonumber \\
 \phi(t)&=& -\int_t^T dt' \Delta(t').
\eea The second line is always solvable by $\Delta(t)=\dot{\phi}(t)$ up to a constant global phase shift provided $\phi(t)$ is everywhere differentiable (smooth). We now focus on the first line. Taking the natural logarithm we arrive at an expression

\bea\label{conditionPacket}
2\Log[A(t)] - \Log[\kappa(t)] = -\int_t^T dt' \kappa(t').
\eea Taking the time derivative of both sides, we arrive at a Bernoulli differential equation \cite{Zwilinger1994}

\bea\label{conditionBernouli}
\kappa^{-2}(t) \frac{d \kappa(t)}{dt} - \frac{2}{A(t)\kappa(t)} \frac{dA(t)}{dt} = -1.
\eea Provided $\frac{1}{A(t)} \frac{dA(t)}{dt}$ is continuous, this is solved by

\bea\label{BernSol}
\kappa(t) = \frac{A^2(t)}{1-\int_{t}^{T} A^2(t') dt' } %where $A(T)$ and $\kappa(T)$ are the wavepacket amplitude and coupling at $t=T$ (the time of detection). Here $\kappa(T)$ is determined by plugging (\ref{BernSol}) into (\ref{conditionPacket}) once a specific form of $A(t)$ is specified.
\eea  Here, $\kappa(t)$ is given by the square of the electromagnetic field, divided by a correction factor accounting for the finite response time imposed by $\kappa(t)$ itself \footnote{We observe from (\ref{BernSol}) that now $A(t)=0 \iff \kappa(t)=0$ for smooth wavepackets, whereas for a general retrodictive distribution (\ref{timeCSolPack}) we find $\kappa(t)=0 \implies A(t)=0$; to generate a smooth wavepacket, $\kappa(t)$ must go to $0$ in the distant past.}. From (\ref{BernSol}), we observe that the only condition imposed on $A(t)$ is that $A^2(t)$ have an antiderivative. We simply require $A^2(t)$ be continuous, which in turn requires $A(t)$ to be continuous. Thus, any wavepacket with smooth phase profile $\phi(t)$ and smooth amplitude $A(t)$ is projected onto by \emph{some} physically realizable single photon detection scheme. \qed

\emph{Special Case: Minimum-Uncertainty Measurement} 

A minimum-uncertainty simultaneous measurement of time and frequency is achieved with a Gaussian time-frequency distribution. We want a temporal wavepacket $\Psi^*(t)$ that is the complex square root a Gaussian distribution 

\bea\label{gaussianWavePacket}
\Psi^*(t)= \frac{1}{(2\pi\sigma^2)^{\frac{1}{4}}} e^{\frac{-(t-t_0)^2}{4\sigma^2}} e^{i\frac{\omega_0}{2}t}
\eea where $\sigma$ is the temporal half-width, and $t_0$ and $\omega_0$ are the central time frequency of the Gaussian distribution. We find that this wavepacket is projected onto by a time-dependent system with constant resonance $\Delta(t) = \omega_0$ and a time-dependent coupling 

\bea\label{ErrfSol}
\kappa(t) = \frac{e^{-\frac{(t-t_0)^2}{2\sigma^2}}}{\sqrt{2\pi\sigma^2}(1+\frac{1}{2}{\rm Erf}[\frac{t_0-T}{\sqrt{2}\sigma}] - \frac{1}{2}{\rm Erf}[\frac{t_0-t}{\sqrt{2}\sigma}])}
\eea as in Fig. \ref{dataCouplingSimp}. Note that the coupling $\kappa(t)$ is dependent on the time of detection $T$ even though the projected state (\ref{gaussianWavePacket}) is $T$-independent, in agreement with the general case (\ref{BernSol}). 

%	\caption{a. Time-dependent amplitude $A(t)$ of a minimum uncertainty (Gaussian) wavepacket in the rotating frame (that is, without the fast oscillations at central frequency $\omega_0$) with time measured w.r.t.  time $t_0$. b. The time-dependent coupling $\kappa(t)$ defined in (\ref{ErrfSol}) that generates this wave-packet. In the time-reversed picture ($t\rightarrow T-t$), this implements a Heisenberg-limited measurement of time and frequency. Here we set time of detection $T=7\sigma$ so that the inefficiency induced by a finite detection time is negligble ($< 10^{-10}$).}

\section{Realistic Measurements Projecting onto Arbitrary Single-Photon States}

The model of a SPD as an isolated two-level system is highly idealized. In a more realistic system, photo detection is an extended process wherein a photon is transmitted into the detector, interacting with the system and triggering a macroscopic change of the photo detector state (amplification) which can then be measured classically. Many theories of single photon detection have been developed over the past century, \cite{glauber1963,kelley1964,scully1969,yurke1984,MandelWolf95,ueda1999,schuster2005,helmer2009,clerk2010,young2018,dowling2018,leonard2019} and indeed there are numerous implementations of SPD technology \cite{Allen39,mcintyer81,marsili2013,frog,wollman2017uv}. Across all systems, we identify these three stages of transmission, amplification, and measurement as universal. In this section, we derive a POVM that incorporates all three stages quantum mechanically, at the end of the section extending the model to include fluctuations of system parameters. The time-dependent two-level system from the previous section enabling arbitrary wavepacket projection is incorporated into the three-stage model as the trigger for the amplification mechanism. We will assume in this analysis that the system is left on for a sufficient time such that the subnormalization of $\Psi(t)$ is minimal and ${\cal W}\approx 1$ (\ref{calW}).

  \begin{figure}[t] % "ht" = here or top
	\includegraphics[width=\linewidth]{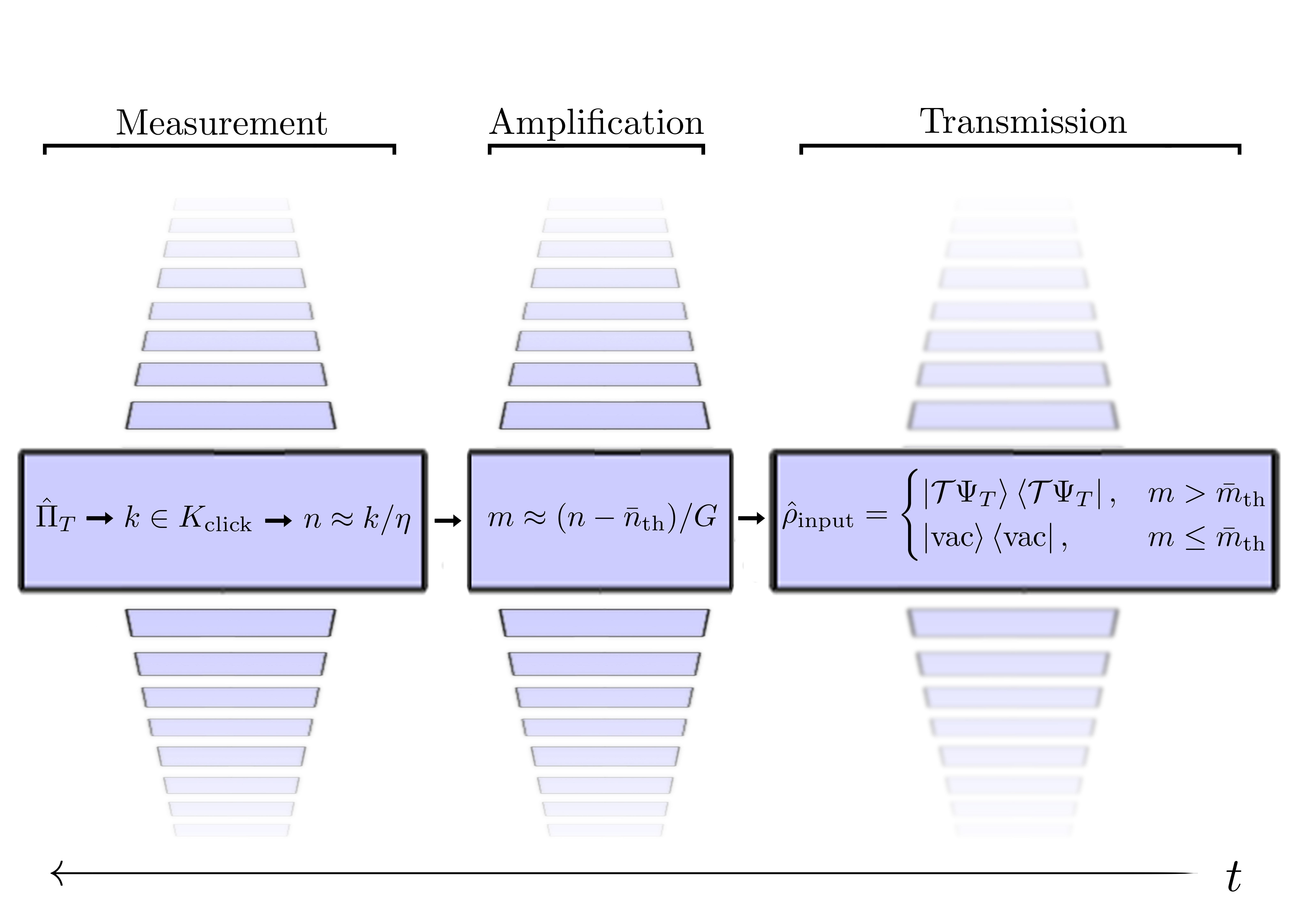} % For Mac OS X PDFs (else append .eps)
	\caption{A POVM description of the three-stage model of photo detection, where the chain of inference (left to right) moves opposite the arrow of time, connecting a macroscopic ``click'' outcome to the state of the input field. A ``click'' outcome represented by the POVM element $\hat{\Pi}_T$ indicates $k\in K_{\rm click}$ excitations were measured post amplification with detection efficiency $\eta$. (Here, we will consider $K_{\rm click}$ to be bounded below by a threshold $k_{\rm min}$ such that $K_{\rm click} = [k_{\rm min},\infty)$.) This suggests $n\approx k/\eta$ (and, strictly, $n\geq k$) excitations were present after in the target mode for amplification. In turn, this indicates that $m\approx (n-\bar{n}_{\rm th})/G\leq n$ excitations were likely incident to the amplification process trigger, with $\bar{n}_{\rm th}$ the expected number of thermal excitations already in the amplification target mode. If $m$ is larger than the expected number of thermal excitations in the amplification trigger mode $\bar{m}_{\rm th}$, we conclude that one (or more) input photon of the form $\ket{\mathcal{T}\Psi_T}\bra{\mathcal{T}\Psi_T}$ was likely present. In addition to the states written explicitly, there are other possible states where $k$, $n$, and $m$ deviate from their most likely values. These states (denoted by parallelograms) contribute to the POVM element, as the internal state of the photo detector is in general highly mixed. Nonetheless, in the end a SPD POVM element $\hat{\Pi}_T$ only projects onto the two input states $\ket{\mathcal{T}\Psi_T}\bra{\mathcal{T}\Psi_T}$ and $\ket{\rm vac}\bra{\rm vac}$, remaining relatively pure.}
	\label{timeFlip}
\end{figure}

In the spirit of what a POVM does (connect present outcomes to probabilistic statements about quantum states in the past), we will begin our endeavor at the very end of the photo detection process following Fig. \ref{timeFlip}. Consider a macroscopic measurement performed at time $T$ with a binary response triggered by $k$ excitations measured in the amplified signal \footnote{There is of course latency in any detector, but this is not interesting: one merely shifts $T\rightarrow T+\delta T$ in each step so that the final POVM matches the timing of a detection event to the quantum state projected onto.}. Such a POVM can be written as a projector onto Fock states in the Hilbert space internal to the system
Barnett2002
\bea\label{clickexample}
\hat{\Pi}_{T}=\sum\limits_{k\in K_{\rm click}} \ket{k,\,T}\bra{k,\,T}. 
\eea Here, we have defined an arbitrary set $K_{\rm click}$ that describes how many amplification excitations must be measured to trigger a macroscopic detection event. We will here assume $K_{\rm click} = [k_{\rm min},\infty)$, describing a lower threshold for a photo detection event. At this stage, we can already see that the internal state the POVM projects onto is highly mixed, but this will not directly translate to an impure measurement on the Hilbert space of input photons. Indeed, this is what we would expect; we do not need to know \emph{precisely} the internal state of the photo detector in order to use it to efficiently detect the  presence of a single photon.

The macroscopic measurement performed on the amplified signal will, in general, be inefficient. We model this in a standard way \cite{Barnett2002}, using a beamsplitter with frequency independent transmission amplitude $\sqrt{\eta}$. We can then rewrite the POVM element (\ref{clickexample}) so that it projects onto Fock states in the amplification target mode prior to the measurement

\bea\label{kmeasurement}
\hat{\Pi}_{T}=\sum\limits_{k\in K_{\rm click}} \sum\limits_{n=k}^\infty  \Pr(n|k) \ket{n,\,T} \bra{n,\,T}
\eea where we have defined 

\bea\label{kmeasureprob}
\Pr(k|n)= {n \choose k}  \eta^k (1-\eta)^{n-k}
\eea the probability to detect $k$ excitations given that there were $n$ excitations in the output mode of the amplification process, which is the same as $Pr(n|k)$ (the probability that given $k$ detected excitations $n$ excitations were incident, needed in (\ref{kmeasurement})) in the absence of prior information \footnote{Indeed, we assume flat priors throughout this paper.}. An inefficiency $1-\eta$ affects photo detection by changing which post-amplification Fock states are projected onto: for a larger $1-\eta$ the distribution of post-amplification states $\ket{n,\,T} \bra{n,\,T}$ that contribute to the final outcome $\hat{\Pi}_T$ becomes larger. This increases the overlap between $\hat{\Pi}_T$ and the null outcome $\hat{\Pi}_{0}$, so that it is harder to distinguish between signal and noise (dark counts). 

We now move one step further back in the chain of inference (Fig. \ref{timeFlip}) so the POVM element $\hat{\Pi}_T$ projects onto the number of excitations $m$ input to the amplification trigger. Amplification is a generic feature of photo detection; without a macroscopic change in the internal state of a photo detector, there is no way to correlate detector outcomes with the presence of a single photon \cite{yang2018,yang2019,zubin2020} (that is, without invoking additional single-excitation detectors in an argument \emph{circulus in probando}). There are many interesting methods for implementing amplification \cite{caves1982,imamoglu2002,yurke2004,clerk2010,metelmann2014,yang2018,yang2019,zubin2020}, but the fundamental quantum limit to amplification of any bosonic Fock state is achieved by a \sch picture transformation \cite{proppamp}
\bea\label{simpleAmp}
\ket{m}_{\rm trig} \ket{M}_{\rm res}\ket{N}_{\rm targ} \longmapsto
\ket{m}_{\rm trig} \ket{M-Gm}_{\rm res}\ket{N+Gm}_{\rm targ}\nonumber \\
\eea such that exactly $G$ excitations are transferred from the reservoir mode to the target mode for each excitation in the trigger mode. In using this expression we do impose a restriction that there must be $M>Gn$ excitations in the reservoir mode, but restrictions of this type are to be expected (the energy for amplification must come from somewhere) and we will be most interested in few photons ($n=0,1,2$) in this analysis. In most physical platforms $G$ will fluctuate \footnote{For instance, electron shelving \cite{dehmelt1975} is exactly described by (\ref{simpleAmp}) in the high-Q cavity limit when at most a single excitation is present with the laser acting as reservoir. Here the fluctuations in $G$ will be sub-Poissonian due to photon bunching.}, as will other (classical) system parameters which we will return to at the end of this section. (Exceptions do exist; for Hamiltonians that implement deterministic amplification schemes [with small integer values for $G$] see Ref.~\cite{bjork1998}.) However, even with a definite gain factor $G$ and number of input excitations $m$, we will still not end up with exactly $n=N+Gm$ excitations if the target mode is initially in a thermal state with mean occupation number $\bar{N}$ (as opposed to a Fock state with exactly $N$ excitations). We now assume this, writing the state of the target mode in the Fock basis 
\bea\label{thermalstate}
\hat{\rho}_{\rm targ}^{(th)} = \sum\limits_{N=0}^\infty  P^{\rm th}_{N,\,T}\Proj{N,\,T},
\eea with the probability for $N$ thermal excitations given by
 \bea\label{thermal}
 P^{\rm th}_{N} = \frac{1}{1+\bar{N}}\left(\frac{\bar{N}}{1+\bar{N}}\right)^N;\,\bar{N}=\frac{1}{e^{\frac{\hbar\omega'}{\tau}}-1},
 \eea  where $\omega'$ and $\tau$ are the frequency and the fundamental temperature of the target mode. Assuming the ideal amplification scheme in (\ref{simpleAmp}), we now write the POVM element $\hat{\Pi}_k$ in terms of the number $m$ excitations that trigger the application mechanism
% \left(\sum\limits_{n=k}^{\infty}  {n \choose k}  \eta^k (1-\eta)^{n-k}   \sum\limits_{m=0}^{{ \rm Int}_- [ \frac{n}{G}]}   P^{\rm th}_{n-Gm}  \right) \ket{\phi_{m,T}}\bra{\phi_{m,T}}   &\nonumber\\
%+\left(\sum\limits_{n=G}^{\infty}  {n \choose k}  \eta^k (1-\eta)^{n-k}  P^{\rm th}_{n-G} \right)\ket{\phi_{m,T}}\bra{\phi_{m,T}} &\\
\bea\label{PhiPOVMtraced:1}
&\!\hat{\Pi}_{T}= \sum\limits_{k\in K_{\rm click}} \sum\limits_{n=k}^\infty  \Pr(n|k)\sum\limits_{m=0}^{{ \rm Fl} [ \frac{n}{G}]}  P^{\rm th}_{n-Gm} \ket{\Psi_T}\bra{\Psi_T}^{\otimes m} \nonumber\\
\eea where we define $\ket{\Psi_T}\bra{\Psi_{T}}^{\otimes 0} = \vacl$ and $P^{\rm th}_{N}=0$ for $N<0$ and have introduced the floor function ${ \rm Fl} [ x] = {\rm Max} [n \in \mathbb{Z} \,|\, n\leq x]$. We can now see the benefit of having a large gain factor G; it shifts the probability distribution over $n$ that corresponds to non-zero excitations in the trigger mode, minimizing its overlap with the probability distribution for zero excitations. In this way, one can dramatically reduce the background noise (dark counts) without decreasing signal by changing ${\rm Min} [K_{\rm click}]$. In (\ref{PhiPOVMtraced:1}) we have reintroduced the state $\ket{\Psi_T}$ defined in (\ref{simpleState}) as the state described by projected onto by the trigger mechanism. As we did in the previous section, we will assume a time-dependent resonance frequency $\Delta(t)$ and decay rate $\kappa(t)$ so that arbitrary pulse-shaping is possible. 

The POVM element in (\ref{PhiPOVMtraced:1}) now projects onto quantum states \emph{internal} to the photo detector. We need to connect the internal continuum of states coupled to the amplification trigger to the external continuum containing the photons we wish to detect (the transmission stage in Fig. \ref{timeFlip}). This is accomplished by introducing an arbitrary two-sided quantum network \cite{proppnet}. This is completely described by a single complex frequency-dependent transmission coefficient $\mathcal{T}(\omega)$ (related to a reflection coefficient at each frequency via $|\mathcal{T}(\omega)|^2 + |\mathcal{R}(\omega)|^2 = 1$ and $\mathcal{R}(\omega)\mathcal{T}^*(\omega) + \mathcal{R}^*\omega)T(\omega) = 0$). We now invoke the single-photon assumption so that there is at most a single excitation input to the quantum network. Any other excitations present in the internal continua will be from internally-generated thermal fluctuations reflected by the quantum network back to the trigger mechanism. In this way, we can construct a POVM element that projects onto product-states $\ket{\psi_{\rm ex}}\bra{\psi_{\rm ex}}\otimes\ket{\psi_{\rm in}}\bra{\psi_{\rm in}}$ of the external and internal continua

%\bea\label{PhiPOVMprod}
%\hat{\Pi}_{T}&=& \sum\limits_{k\in K_{\rm click}} \sum\limits_{n=k}^\infty  \Pr(n|k) \left( P^{\rm th}_{n} \vacl\otimes \vacl \phantom{\sum\limits_{m=1}^{{ \rm Int}_- [ \frac{n}{G}]} } \right.\hfill \nonumber \\ 
%&+&\sum\limits_{m=1}^{{ \rm Int}_- [ \frac{n}{G}]}  P^{\rm th}_{n-Gm} r^{2m}\vacl\otimes\ket{RG_T}\bra{RG_T}^{\otimes m} \\ \nonumber
%&+&\left.\sum\limits_{m=1}^{{ \rm Int}_- [ \frac{n}{G}]}  P^{\rm th}_{n-Gm} m t^2 r^{2(m-1)}\ket{TG_T}\bra{TG_T} \otimes\ket{RG_T}\bra{RG_T}^{\otimes m-1} \right) \nonumber
%\eea

\begin{widetext}\bea\label{PhiPOVMprod}
\hat{\Pi}_{T}&=& \sum\limits_{k\in K_{\rm click}} \sum\limits_{n=k}^\infty  \Pr(n|k) \left( P^{\rm th}_{n} \vacl\otimes \vacl  + \sum\limits_{m=1}^{{ \rm Fl}[ \frac{n}{G}]}  P^{\rm th}_{n-Gm} \beta^{2m}\vacl\otimes\ket{\mathcal{R}\Psi_T}\bra{\mathcal{R}\Psi_T}^{\otimes m}\phantom{\sum\limits_{m=1}^{{ \rm Fl} [ \frac{n}{G}]} } \right. \\ \nonumber
&+&\left.\sum\limits_{m=1}^{{ \rm Fl} [ \frac{n}{G}]}  m P^{\rm th}_{n-Gm} \alpha^2 \beta^{2(m-1)}\ket{\mathcal{T}\Psi_T}\bra{\mathcal{T}\Psi_T} \otimes\ket{\mathcal{R}\Psi_T}\bra{\mathcal{R}\Psi_T}^{\otimes m-1} \right). \nonumber
\eea\end{widetext} The first term corresponds to dark counts generated from thermal excitations post-amplification and the second term corresponds to dark counts generated by thermal excitations that then trigger the amplification mechanism. Only the third term contains a projection onto a photon to be detected. (The multiplicative factor $m$ in the third term is combinatorial in origin: $m$ total excitations in the trigger mode with $m-1$ generated from thermal fluctuations.) In writing (\ref{PhiPOVMprod}) we have defined transmitted and reflected normalized single-photon states and coefficients 

\bea\label{singlephotonstate}
\ket{\mathcal{T}\Psi_T}&=&\frac{1}{\alpha} \int_{-\infty}^\infty\,d\omega \tilde{\Psi}(\omega)\mathcal{T}^*(\omega)e^{i\,\omega\,T}\hat{a}^{\dagger}(\omega)\ket{\textnormal{vac}} \nonumber \\
\ket{\mathcal{R}\Psi_T}&=&\frac{1}{\beta} \int_{-\infty}^\infty\,d\omega \tilde{\Psi}(\omega)\mathcal{R}^*(\omega)e^{i\,\omega\,T}\hat{b}^{\dagger}(\omega)\ket{\textnormal{vac}}\nonumber\\
\alpha&=&\sqrt{\int d\omega |\tilde{\Psi}(\omega)|^2  |\mathcal{T}(\omega)|^2}\nonumber\\
\beta&=&\sqrt{\int d\omega |\tilde{\Psi}(\omega)|^2  |\mathcal{R}(\omega)|^2} 
\eea where $\hat{a}^{\dagger}$ and  $\hat{b}^{\dagger}$ are the creation operators for the external and internal continua and we have defined a Fourier-transformed wavepacket for the amplification trigger mode $\tilde{\Psi}(\omega) = {\rm FT}[\sqrt{\kappa(t)}\Psi(t)]$. We can now see how pre-amplification dark counts (the second line of (\ref{PhiPOVMprod})) can be suppressed: by reducing the overlap of $|\tilde{\Psi}(\omega)|^2$ and $|\mathcal{R}(\omega)|^2$, that is, by only amplifying the frequencies we wish to detect so that $\beta^2\ll 1$. In this case, the POVM element (\ref{PhiPOVMprod}) will be dominated by the $m=1$ term of the third line (the signal to be detected with no thermal excitations), as well as potentially the first line. (To reiterate, these are dark counts post-amplification, but these can be reduced by amplifying at a high frequency such that $\hbar\omega'\gg \tau$, where $\omega'$ and $\tau$ are the frequency and fundamental temperature of the target mode.)

%We assume the internal continuum is in a thermal state 
%\bea\label{thermalstate}
%\hat{\rho}_{\rm int}^{(th)}= \sum\limits_{j=0}^\infty  P^{'\rm th}_{j}\Proj{j}
%\eea where the Fock states do not occupy a monochromatic mode (as they did for the target mode) but instead the reflected wavepacket defined above in (\ref{singlephotonstate}), so that the probability to have $j$ excitations is
%\bea\label{thermal}
% \!\!\!P^{'\rm th}_{j} = \frac{1}{1+\bar{j}}\left(\frac{\bar{j}}{1+\bar{j}}\right)^j;\,\bar{j}=\int d\omega \frac{|\tilde{\Psi}(\omega)|^2 |R(\omega)|^2}{r^2\left(e^{\frac{\hbar\omega}{k_b T'}}-1\right)}
% \eea 
% where $T'$ is the temperature of this internal continua \footnote{The authors apologize for the overuse of the letter ``T'' in this paper.}. 
 
 Finally, we trace over the internal continua, which we assume is in a thermal state with fundamental temperature $\tau'$  so the POVM projects onto the external continua only
 
\bea\label{POVMExt}
\hat{\Pi}_{T}&=& \sum\limits_{k\in K_{\rm click}} \sum\limits_{n=k}^\infty  \Pr(n|k) \left(\sum\limits_{m=0}^{{ \rm Fl} [ \frac{n}{G}]}  P^{\rm th}_{n-Gm} P^{'\rm th}_{m} \beta^{2m} \right.\vacl \nonumber \\
&+& \left. \sum\limits_{m=1}^{{ \rm Fl} [ \frac{n}{G}]}  m P^{\rm th}_{n-Gm} P^{'\rm th}_{m-1} \alpha^2 \beta^{2(m-1)}\ket{\mathcal{T}\Psi_T}\bra{\mathcal{T}\Psi_T} \right) \nonumber \\ 
&\equiv& w_0 \vacl + w_T \ket{\mathcal{T}\Psi_T}\bra{\mathcal{T}\Psi_T}
\eea where in the last line we have absorbed the sums in front of the two projectors into weights so that the POVM element has the form of (\ref{POVM1}) and with $P^{'\rm th}_{j} $ the probability to have $j$ thermal excitations (now in the non-monochromatic reflected mode defined in (\ref{singlephotonstate})). For a finite detector on-time $T_0> -\infty$, the weights $w_0$ and $w_T$ will be slightly less than in (\ref{POVMExt}) due to wavepacket sub-normalization (\ref{calW}). However, this deviation is negligible provided the detector is left on for a time comparable to the temporal mode's width.

We now reconsider the question of projecting onto an arbitrary wavepacket, including the full quantum description. We find that this is possible to do in principle, \emph{provisio} $\mathcal{T}(\omega)$ is nowhere zero (except at infinity), a result we will now prove. That is, we can ensure that the single-photon wavepacket $\ket{\mathcal{T}\Psi_T}$ has any desired (smooth) shape and will be projected onto with a high-efficiency and high-purity measurement.

\lettersection{Proof}Consider a photon with complex normalized spectral wavepacket $\tilde{f}(\omega)$. If detection is achieved with a time-dependent two-level system preceded by a quantum network with filtering transmission function $\mathcal{T}(\omega)$, the system will project onto a state $\ket{\mathcal{T}\Psi_T}$ as defined in (\ref{singlephotonstate}). In the low-noise limit this will be the only state projected onto by the (pure) POVM element. From the Born rule, the probability of detection will be 
\bea\label{ProbDetectProofT}
P_T = w_T \frac{1}{\alpha^2} \left | \int_{-\infty}^{\infty} d\omega \tilde{f}(\omega) \tilde{\Psi}(\omega) \mathcal{T}^*(\omega) \right |^2
\eea with $w_T$ the overall weight given by (\ref{POVMExt}) and maximum possible detection efficiency, which can be arbitrarily close to unity. It is possible to achieve $P_T = w_T$ (mode-matched detection) in (\ref{ProbDetectProofT}) if and only if
\bea\label{ProbDetectProofSol}
\tilde{\Psi}(\omega) =\frac{\tilde{f}^*(\omega)}{\mathcal{T}(\omega)} e^{i\omega T}. 
\eea From (\ref{timeCSolPack}), we know that it is possible to generate an arbitrary temporal wavepacket  ${\rm FT}^{-1}[\tilde{\Psi}(\omega)]=\sqrt{\kappa(t)}\Psi(t)$ from a time-dependent two-level system.  The Fourier transform of a continuous smooth function is itself smooth and continuous. Thus, if the right hand side of (\ref{ProbDetectProofSol}) is a well-defined spectral wavepacket (smooth and continuous), one can find functions $\kappa(t)$ and $\Delta(t)$ such that $\tilde{\Psi}(\omega)$ has the form of  (\ref{ProbDetectProofSol}). \qed

\lettersection{Remark} Arbitrary wavepacket detection (and thus Heisenberg-limited simultaneous measurements of time and frequency) is in principle possible only when there are no photonic band gaps induced by the filter; if $\mathcal{T}(\omega')=0$ for some frequency $\omega'$, there is simply no way to compensate for the lost information about $\omega'$. Photonic band gaps are a generic feature of parallel (and hybrid) quantum networks \cite{proppnet} as well as certain non-Markovian systems \cite{Garraway2006}. Network/reservoir engineering must be employed to ensure any $\omega'$ where $T(\omega')=0$ is not a frequency of interest. %However, if $\tilde{G}(\omega')=0$ as well there is no issue, there is no need to recover information about that frequency in the first place. 

The POVM $\{\hat{\Pi}_{T},\,\hat{\Pi}_0\}$ with $\hat{\Pi}_{T}$ defined in (\ref{POVMExt}) and $\hat{\Pi}_{0}=\idh-\hat{\Pi}_{T}$ provides a complete description of the single photon detection process that is fully quantum from beginning to end (Fig. \ref{timeFlip}). However, there is a final element that must be considered to make the description applicable to laboratory systems: classical parameter fluctuations. For continuous parameter fluctuations over any system parameter or set of system parameters $X$, these are naturally incorporated

\begin{center}
\begin{figure}[t]
%\begin{tabular}{| c | c | c|}
%\hline
%Symbol& System Parameter & Fluctuations \\
%\hline\hline
%$k_{\rm min}$& Macroscopic Detection Threshold & $w_0,\,w_T$\\
%\hline
%$\eta $& Macroscopic Detection Efficiency & $w_0,\,w_T$\\
%\hline
%$\tau$& Target Mode Temperature & $w_0,\,w_T$\\
%\hline
%$\omega' $& Target Mode frequency & $w_0,\,w_T$\\
%\hline
%$G$& Amplification Gain & $w_0,\,w_T$\\
%\hline
%$\tilde{\Psi}(\omega) $& Amplification Trigger Mode & $w_0,\,w_T,\, \ket{\mathcal{T}\Psi_T}$\\
%%\hline
%%$T $& Detection Time & $\ket{TG_T}$\\
%\hline
%$\tau' $& Internal Continua Temperature & $w_0,\,w_T$\\
%\hline
%$\mathcal{T}(\omega)$& Transmission Coefficient & $w_0,\,w_T\, \ket{\mathcal{T}\Psi_T}$\\
%\hline
%\end{tabular}
	\includegraphics[width=\linewidth]{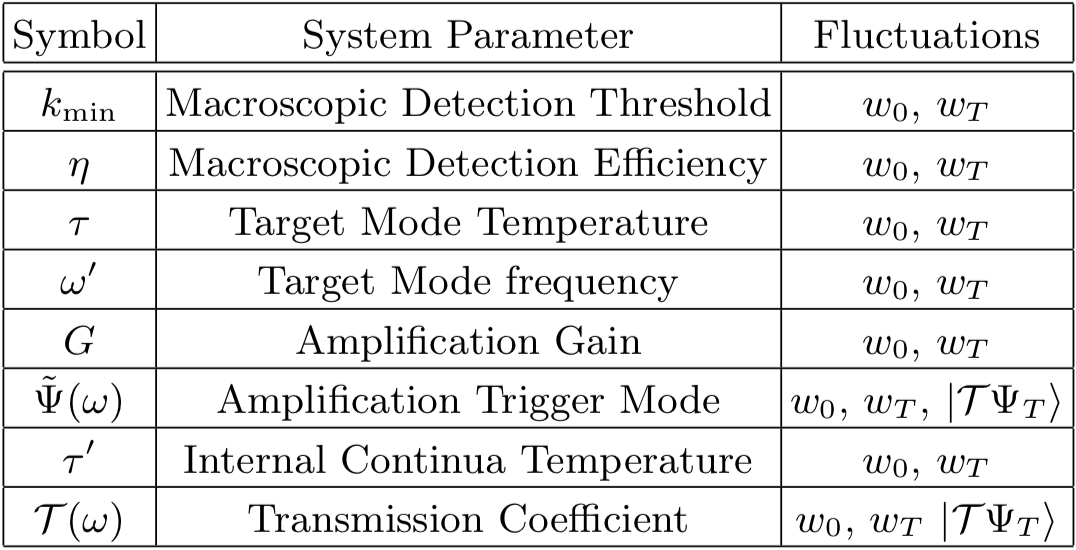} % For Mac OS X PDFs (else append .eps)
\caption{The effects of classical fluctuations in system parameters on the final POVM: either the weights and states are changed, or only the weights are changed. The fluctuations over the functions $\tilde{\Psi}(\omega) $ and $\mathcal{T}(\omega)$ (and thus $\mathcal{R}(\omega)$ by unitarity) could be caused by fluctuations in other system parameters (decay rates, resonances) internal to those functions. A subset of fluctuations in $\tilde{\Psi}(\omega) $ are fluctuations in the time of detection $T$. Importantly, these shift the wavepacket $\Psi(t)$ projected onto, resulting in a mixed measurement with larger temporal uncertainty (jitter) that depends on the ratio of the fluctuations in $T$ to the width of the temporal wavepacket.}\label{table}
\end{figure}
\end{center}

\bea\label{POVMFluct}
\!\hat{\Pi}_{T} = \int dX {\rm Pr}(X) \left(w_0 \vacl + w_T \ket{\mathcal{T}\Psi_T}\bra{\mathcal{T}\Psi_T}\right)\nonumber\\
\eea where we have assumed a (known) probability distribution $ {\rm Pr}(X)$. In (\ref{POVMFluct}), the system parameter(s) $X$ could be such that only the weights $w_0$ and $w_T$ depend on $X$, or $X$ could be such that the state $ \ket{\mathcal{T}\Psi_T}$ depends on $X$ as well (for a summary, see Fig. \ref{table}). In the case of the latter, the POVM will become less pure and will need rediagonalization to determine which states are projected onto \footnote{By varying certain key parameters, it is possible to induce an exceptional-point structure in (\ref{POVMFluct}), for instance, by introducing a discrete probability distribution over resonance frequencies corresponding to classical ignorance about a discrete set of detector settings. Here the exceptional point occurs when the frequencies are made degenerate, which (since the discrete states have identical quantum numbers) is forbidden by unitarity. Here the range over the resonances are distributed is the exceptional-point parameter \cite{Heiss2012}.}. This final POVM not only includes ignorance about the internal state of the photo detector as was depicted in Fig. \ref{timeFlip}, but also classical ignorance about the state of the photo detector due to system-lab interactions. 

For example, let us start with our ideal detector from Section II, which projects onto a Gaussian wave packet (\ref{gaussianWavePacket}) with a central time $t_0$ a fixed duration before $T$ and a width $\sigma$ determined by the specific form of $\kappa(t)$.   Suppose the parameter $T$ fluctuates such that $t_0$ fluctuates but $\sigma$ stays fixed. For definiteness, assume a Gaussian distribution for the central time $\tau_0$ of the wave packet
\be
{\rm Pr}(\tau_0)=\frac{1}{\sqrt{2\pi w^2}}\exp(-(\tau_0-t_0)^2/2w^2)
\ee
with $w$ the width of that distribution.
One effect on the POVM ``click'' element of this uncertainty about the value of $\tau_0$ is that its purity decreases.
Since that consequence has been discussed in a slightly different context in Ref.~\cite{spectralPOVM}, we focus here on a second effect: that the probability of detecting the single-photon wave packet (\ref{gaussianWavePacket}) our detector was designed to detect perfectly decreases.
The probability to detect our favorite wave packet given a parameter $\tau_0$ is 
\be
{\rm Pr}(t_0|\tau_0)=\exp(-(\tau_0-t_0)^2/4\sigma^2).
\ee 
Averaging this probability over the Gaussian distribution over $\tau_0$ gives
\be
\int\! d\tau_0\, {\rm Pr}(\tau_0){\rm Pr}(t_0|\tau_0)=\frac{1}{\sqrt{1+(w/\sigma)^2/2}},
\ee
which in the limit of $w\rightarrow 0$ reaches 1, as it should, and which for large $w$ decays to zero as $\sqrt{2}\sigma/w$. If one additionally includes other realistic factors (finite temperature, finite gain, and the threshold value $k_{\rm min}$ of $K_{\rm click} = [k_{\rm min},\infty)$), the conditional probability ${\rm Pr}(t_0|\tau_0)$ remains uniform across the parameter $\tau_0$ and thus the overall detection efficiency decreases by the factor $w_T$ as defined in (\ref{POVMExt}).

We may take this description one step further and 
consider $X$ not as  a set of classical parameters but as (discrete) outcomes $X_n$ of quantum measurements.
That is, there is a higher-level POVM $\pi_n$ describing measurements on one or more auxiliary system (e.g., a quantum clock for measuring time \cite{maccone2020}), with the possible outcomes labeled by the parameters $X_n$.
When the probability of the measurement outcome $X_n$ is ${\rm Pr}(X_n)$, then we would get the same type of mixed POVM (\ref{POVMFluct})---but with the integral replaced by a sum over $n$---when different outcomes correspond to orthogonal outcomes of a standard Von Neumann measurement, such that ${\rm Tr}(\pi_n \pi_m)=\delta_{nm}$.
In case different outcomes are not orthogonal (as would be the case for quantum measurements of phase and time), 
the probability distribution ${\rm Pr}(X)$
must be replaced according to
\be
\int\! dX \,{\rm Pr}(X) \hat{\Pi(X)} \longmapsto \sum_n \sum_m
\frac{{\rm Tr} (\pi_n \pi_m)}{{\rm Tr} (\pi_m)}{\rm Pr}(X_m) \hat{\Pi}(X_n),
\ee
since an outcome $\pi_m\neq \pi_n$ may still project onto the single-photon state corresponding to $X_n$.
Apart from this change the result (\ref{POVMFluct})  retains the same form.

\section{Applications}

\begin{figure}[t] % "ht" = here or top
	\includegraphics[width=1\linewidth]{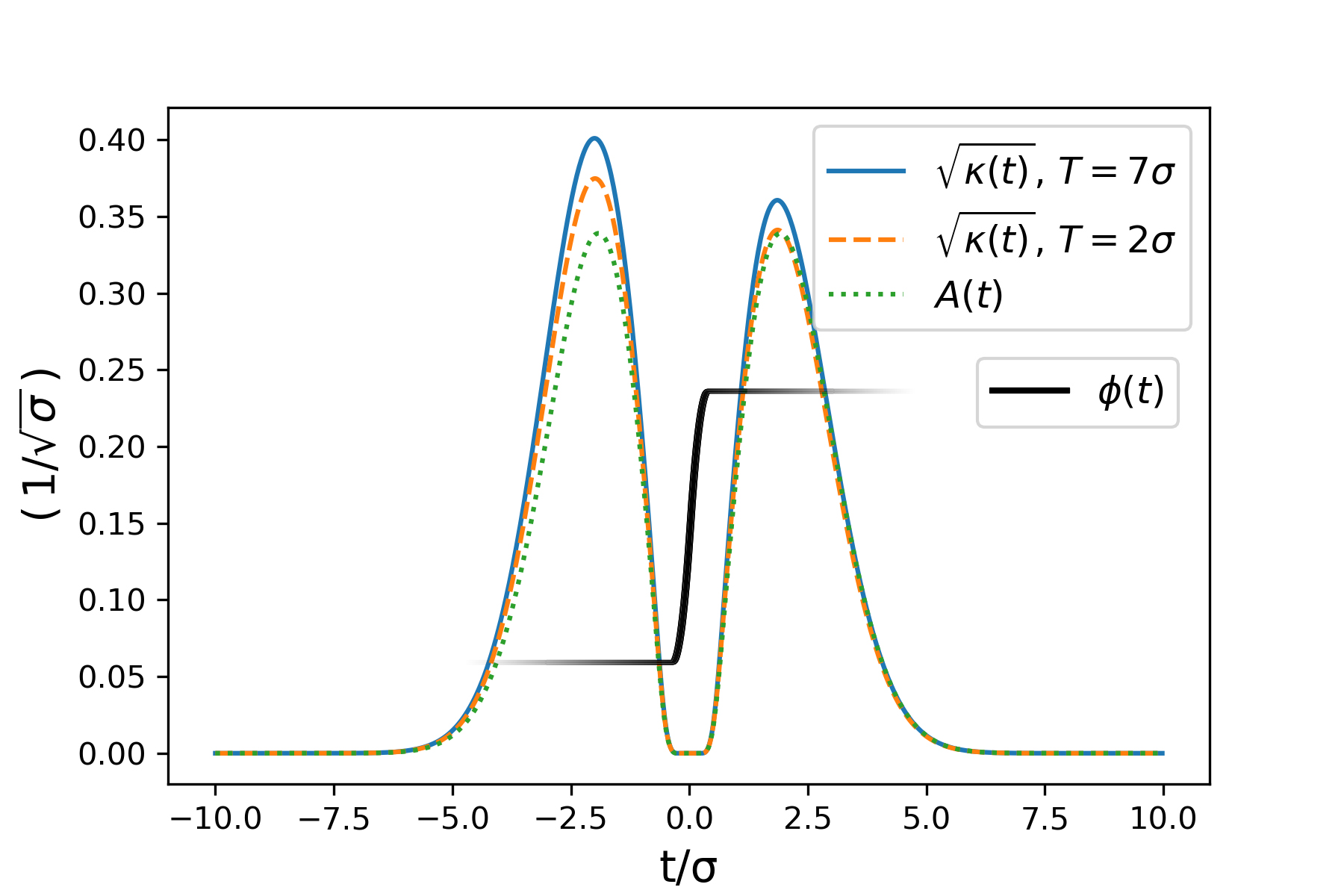} % For Mac OS X PDFs (else append .eps)
	\caption{The time-dependent coupling $\kappa(t)$ that generates a wavepacket exactly orthogonal to the minimum uncertainty Gaussian wavepacket in Fig. \ref{dataCouplingSimp} are plotted in units of $1/\sqrt{\sigma}$ for times of detection $T=7\sigma$ (solid blue line, ${\cal W} = 1-10^{-5}$) and $T=2\sigma$ (dashed orange line, ${\cal W} = 0.72$) and detector on-time $T_0=-\infty$. In the rotating frame, the wavepacket's time-dependent amplitude $A(t)$ (green dotted line, also in units of $1/\sqrt{\sigma}$) is an approximate first-order Hermite-Gaussian pulse, where the singular region of zero has been expanded with half-width of $z=0.5\sigma$. In this way, the wavepacket's phase (thick black line, not to scale) can go from $0$ to $\pi$ in a finite time, whereas the Heaviside phase-flip in the exact Hermite-Gaussian pulse requires an unphysical delta-function detuning (second line of (\ref{sepExpress}). Both the phase and amplitude have been convolved with a triangular smoothing function with full-width $s=0.5\sigma$, ensuring $A(t)$ is continuously differentiable to first order \cite{Boehme1966,deBoor1972}, which is pre-requisite for solving the Bernoulli equation (\ref{conditionBernouli}). For any finite $z\geq s>0$ the phase flip [here implemented with a triangular detuning $\Delta(t)$] occurs while the amplitude is zero. This ensures exact orthogonality of the approximate first-order Hermite-Gaussian to the gaussian pulse. In the limit $s,z\rightarrow 0$, an exact first-order Hermite-Gaussian is recovered. This smoothing procedure generalizes to higher order Hermite-Gaussian pulses, forming a mutually unbiased basis for efficient detection of higher dimensional qudits \cite{reddy2015}.}
	\label{OrthogonalState}
\end{figure}

Using the time-dependent two-level system, we are able to project onto orthogonal quantum states (Fig. \ref{OrthogonalState}). This enables efficient detection of photonic qubits, an essential component of any quantum internet \cite{kimble2008,lukens2017}. More generally, temporal modes provide a complete framework for quantum information science \cite{reddy2015}, with efficient detection of orthogonal modes (and their superpositions to create mutually unbiased bases) a key ingredient. Fully manipulable temporal modes also play a key role in error-corrected quantum transduction \cite{vanenk1997}, where a time-reversed temporal mode can restore an unknown superposition in a qubit. Here, efficient detection of arbitrary temporal modes is essential so that quantum jumps out of the dark state are efficiently heralded. 

High-purity measurements that project onto orthogonal single-photon wavepackets also enable super-resolved measurements \cite{bonaszek2017}. Suppose we have two single-photon sources emitting almost identical pure states differing slightly in either emission time or central frequency
\bea
\ket{\tilde{\phi}_1} &=\frac{\ket{\phi_1} + \sqrt{\epsilon} \ket{\phi_2}}{\sqrt{1+\epsilon}}\nonumber\\
\ket{\tilde{\phi}_2} &=\frac{\ket{\phi_1} -\sqrt{\epsilon} \ket{\phi_2}}{\sqrt{1+\epsilon}}
\label{superres}
\eea with $\braket{\tilde{\phi}_1}{\tilde{\phi}_2}$ real, $\epsilon\ll 1$, and $\braket{\phi_1}{\phi_2}=0$.
Alternatively,  we may imagine a single source of light but the light we receive may have either been slightly Doppler-shifted or it may have been slightly delayed.

 Suppose now that we receive one photon that could equally likely be from either source so that our input state is
\bea
\hat{\rho} = \frac{1}{2} \ket{\tilde{\phi}_1}\bra{\tilde{\phi}_1} +  \frac{1}{2}\ket{\tilde{\phi}_2}\bra{\tilde{\phi}_2}. \label{inputstatemixed}
\eea If we can measure both $\hat{\Pi}_1=\eta \ket{\phi_1}\bra{\phi_1}$ and $\hat{\Pi}_2=\eta \ket{\phi_2}\bra{\phi_2}$ (that is, if we have separate photodetectors with these (pure) POVM elements, or a single non-binary-outcome photo detector), then we find the probability of clicks
\bea
P_1 &=& {\rm Tr} \left[ \hat{\Pi}_1 \hat{\rho} \right] = \eta \frac{1}{1+\epsilon} \nonumber\\
P_2 &=& {\rm Tr} \left[ \hat{\Pi}_2 \hat{\rho} \right] = \eta \frac{\epsilon}{1+\epsilon}
\label{probclicks}
\eea so that the ratio of clicks gives a direct estimate of $\epsilon$, even for low efficiency $\eta$. Here all that is needed for time-frequency domain super-resolved measurement of $\epsilon$ are SPDs with time-dependent couplings and resonance frequencies as opposed to nonlinear optics \cite{donohue2019}.

In traditional quantum key distribution (QKD) schemes (that is, \emph{not} measurement device independent (MDI)-QKD), specification of the measurement POVM is essential to robust security proofs \cite{norbert2004,qi2006,qi2009}. Here, we have verified several assumptions about an eves-dropper's capabilities common in security proofs: that high-purity measurements are possible, that high efficiency measurements are possible, and (for continuous-variable (CV)-QKD proofs) minimum time-frequency uncertainty measurements are possible. In particular for CV-QKD, an eavesdropper can perform measurements that project onto variable-width spectral modes, disrupting temporal correlations between Alice and Bob (who are assumed to use fixed time-frequency bins) \cite{bourassa2019}. Here, the capacity to adjust the width of the spectral mode $\tilde{\Psi}(\omega)$ provides Alice and Bob a new strategy to mitigate Eve's attack and extract a secure key. 

More generally, detector tomography is an important tool across implementations of single-photon and number-resolved photo detection \cite{dariano2004,lundeen2009,coldenstrodt2009,Ma2016}. Real-time tomography could be useful in QKD protocols resistant to ``trojan-horse attacks'' \cite{gisin2006} or any SPD platform subject to time-dependent environmental parameter fluctuations: for instance, atmospheric turbulence in MDI-QKD \cite{Hu2018} or interplanetary medium in deep space classical communications \cite{Banaszek2019}. Recently tomography speed-ups have been achieved using machine learning assisted tomography protocols \cite{abj2019}. The POVMs derived in this paper provide priors which can further speed up detector tomography \cite{Heinosaari2013}. These include approximate effects of environmental fluctuations as outlined in Fig. \ref{table} and a global optimum POVM for single photon detection (\ref{POVMExt}) which can be used to incorporate detector calibration and optimization into \emph{in situ} tomographic protocols. 

\section{Conclusions}

Having gone through applications of our work, we return to the fundamental (as opposed to practial) limits to single photon detection and their implications, as well as possible experimental implementations.

Here we have constructed single-photon measurements that are Heisenberg-limited in two ways: the first is that they can project onto Gaussian time-frequency states as illustrated in Fig. \ref{dataCouplingSimp}, and the second is that the amplification scheme reaches a Heisenberg-limited (linear in the gain $G$) signal-to-noise ratio, surpassing the standard quantum limit (a signal-to-noise ratio going like of $\sqrt{G}$) \cite{proppamp}. Achieving these simultaneously is possible in principle with no drawback. Indeed, the only stringent tradeoff we encounter in this analysis is between efficiency and photon counting rate, which becomes substantial when an SPD is reset at a faster rate than $\sim 1/\Delta t$. (The photon does not have sufficient time to excite the two-level system with high probability before the system is reset.) For other figures of merit, we find that they are either independent, or deteriorate together \footnote{For instance, inefficiency and dark counts both increase with the coefficient $\rho$ in (\ref{singlephotonstate}) when one considers an amplification scheme like electron shelving, where the absorption of one excitation precludes the absorption of a second.}. While it does appear from (\ref{POVMExt}) that improving efficiency also increases dark counts, these are decoupled by ensuring the coefficient $\beta\ll 1$---that is, by making $\mathcal{T}(\omega)$ broader than $\tilde{\Psi}(\omega)$ as in (\ref{singlephotonstate}). While it is commonsense that one should only amplify the frequencies they wish to detect, our work clarifies how enormously important this is. The dark counts produced in this way are insuperable; they cannot be removed post-amplification without removing the single-photon signal as well.

Another conclusion from this work is rather optimistic. Here we have given a quantum description of an entire single photon detection process projecting onto arbitrary single photon states and the only fundamental limitations encountered are Heisenberg limits. Incorporating realistic descriptions of amplification and a final measurement reduce efficiency and increase dark counts, but even so a Heisenberg-limited measurement is still  achievable in principle. Similarly, incorporating the filtering of a first irreversible step does not impede implementation of Heisenberg-limited measurements provided no frequencies are completely blocked from entering the trigger mechanism.  Even considering  parameter fluctuations (\ref{POVMFluct}) in internal temperatures $\tau$ and $\tau'$, amplification frequency $\omega'$, and amplification gain factor $G$---which are unavoidable in any realistic system---Heisenberg limited time-frequency measurements are achieved. To the authors' knowledge, this is first proposed quantum procedure for reaching Heisenberg limited time-frequency measurements in a realistic quantum system. In addition to being a fundamental limit to SPD performance, probing Heisenberg limits paves the way for future experimental tests of foundational quantum theory.

Experimental implementation of the POVM derived in this work relies on several well-established technological elements. The first is passive filtering to couple the photo detecting system to an external continuum of states (i.e. transmission). This could be an optical fiber or a lens to focus light (e.g. onto a set of molecules as in the eye). Additionally, an electron shelving three-level system turned on at a time $T$ is needed to implement amplification of the input single photon into a macroscopic signal with minimal noise (see Ref.~\cite{dehmelt1975} for details on electron shelving). The last stage of the photo detecting process is an inefficient measurement of the macroscopic signal, which can be implemented with an avalanche photo diode or a photo multiplier tube. In this work, the three stages of photo detection (transmission, amplification, and measurement) have been considered separate and sequential so that we could elucidate the fundamental limits and trade-offs that arise at each step. However, there has been recent progress unifying these key ingredients for photo detection in a single Hamiltonian \cite{Biswas2020}. Future work will connect that progress to the limits derived in this paper, and elucidate how the shift from a discrete photo detection event at time $T$ to a continuously monitored system affects the final photo detection POVM.

\vspace{1em}

This work is supported by funding from
DARPA under
Contract No. W911NF-17-1-0267, as well as by National Science Foundation Grant No. PHY-1630114.
\bibliography{ThePOVM}

\end{document}